\documentclass[pra,twocolumn,showpacs,superscriptaddress,floatfix]{revtex4}
\usepackage[dvipdfmx,dvipdfmx]{graphicx}
\usepackage{color}
\usepackage{subfigure}
\usepackage{epsfig}
\usepackage{float}
\usepackage{epstopdf}

\usepackage{amsmath}	

\newcommand{\ket}[1]{\left | #1 \right \rangle}
\def\k(#1){|#1\rangle}
\newcommand{\bra}[1]{\left \langle #1 \right |}

\newcommand{\beq}{\begin{equation}}
\newcommand{\eeq}{\end{equation}}
\newcommand{\beqa}{\begin{eqnarray}}
\newcommand{\eeqa}{\end{eqnarray}}
\newcommand{\beqan}{\begin{eqnarray*}}
\newcommand{\eeqan}{\end{eqnarray*}}

\newcommand{\affA}{%
\affiliation{
 National Institute of Information and Communications Technology,
 4-2-1 Nukui-kita, Koganei, Tokyo 184-8795, Japan}
     }
\newcommand{\affB}{%
\affiliation{
 Sophia University,
 7-1 Kioicho, Chiyoda-ku, Tokyo 102-8554, Japan}
}

\newcommand{\affC}{%
\affiliation{
Department of Physics, Technical University of Denmark, Building 309, 2800 Lyngby, Denmark}
}

\begin{document}

\title{Projective measurement onto arbitrary superposition of weak coherent state bases}

\author{Shuro Izumi }
\affA \affB \affC
\author{Masahiro Takeoka}%
\affA
\author{Kentaro Wakui}%
\affA
\author{Mikio Fujiwara}%
\affA
\author{Kazuhiro Ema}%
\affB
\author{Masahide Sasaki}%
\affA
%



\begin{abstract}
One of the peculiar features in quantum mechanics is that a superposition 
of macroscopically distinct states can exist. In optical system, 
this is highlighted by a superposition of coherent states (SCS), 
i.e. a superposition of classical states. 
Recently this highly nontrivial quantum state and its variant have 
been demonstrated experimentally. 
Here we demonstrate the superposition of coherent states 
in {\it quantum measurement} which is also a key concept 
in quantum mechanics. 
More precisely, we propose and implement a projection measurement onto 
an arbitrary superposition of two weak coherent states in optical system. 
The measurement operators are reconstructed experimentally 
by a novel quantum detector tomography protocol. 
Our device is realized by combining the displacement operation 
and photon counting, well established technologies, 
and thus has implications in various optical 
quantum information processing applications. 
\end{abstract}

\maketitle

\section{Introduction}
 Quantum measurement plays an essential role in 
Quantum Information Processing (QIP). 
In quantum optical system, the standard measurements are homodyne detector and photon detector that measure the physical quantities of light: quadrature amplitudes and photon numbers, respectively. 

However, one can consider more general quantum measurement that has no correspondence to these standard physical quantities, more precisely, any measurement satisfying the mathematical condition of the positive operator valued measure (POVM) formalism. The example of such non-standard measurement considered here is the projection measurement onto the superposition of coherent states (SCS),
$a_0 |\alpha\rangle + a_1 |{-}\alpha\rangle$, where 
$\ket{\pm \alpha}$ are the coherent state, i.e. classical state, 
with amplitude $\pm\alpha$. 
More precisely, we consider the arbitrary projection measurement 
in the space spanned by the SCS bases $|C_\pm\rangle = 
(|\alpha\rangle \pm |{-}\alpha\rangle)/\mathcal{N}_\pm$ 
($\mathcal{N}_\pm$ are the normalization factors):  
\begin{eqnarray}
\label{eq1}
\left\{ 
\begin{array}{l}
|\pi^{SCS}_0\rangle = c_0 |C_+\rangle + c_1 \mathrm{e}^{i \phi} |C_-\rangle, 
\\ 
|\pi^{SCS}_1\rangle = c_1  \mathrm{e}^{-i \phi} |C_+\rangle - c_0 |C_-\rangle,  
\end{array}
\right.
\end{eqnarray}
where $c_0$ and $c_1$ are real and satisfy $c_0^2+c_1^2=1$ 
and $\phi$ denotes the relative phase between $\ket{C_{\pm}}$.

Each vector of the measurement in Eq.~(\ref{eq1}) is equivalent to the SCS state, which is a typical example of macroscopic quantum superposition (and thus sometimes regarded as ``Schr\"{o}dinger cat state'') showing highly nonclassical properties.
Generation of such optical states have been experimentally accomplished 
by several groups \cite{Ourjoumtsev_Science, Jonas_cat, wakui_cat, 
Ourjoumtsev_nature,Takahashi_cat, jonas_cat2,Gerrits_cat,Yukawa_cat,etesse_cat}
for relatively small $\alpha$. 
Especially in 
Ref.\cite{jonas_cat2}
, generation of the approximate SCS with 
arbitrarily controlled $\{c_0, c_1, \phi\}$ is demonstrated. 
On the other hand,
a few attempts have been made for the exploration of the measurement 
described by the SCS bases. 
It is well known that a specific projection measurement $\ket{C_{\pm}}$ (i.e. $c_0=1$,$c_1=0$,$\phi=0$) is realized by the parity measurement of photon numbers. 
However, the implementation of the measurement 
for general $\{c_0, c_1, \phi\}$ remains as a challenge.

In this paper, 
we propose and experimentally demonstrate physical implementation of 
the SCS measurement with arbitrary $\{c_0, c_1, \phi\}$ in the regime 
of small $\alpha$. 
The structure of the implemented measurement (i.e. its positive operator 
valued measure (POVM)) is reconstructed by the quantum detector tomography 
(QDT) \cite{Lundeen09,Brida,Lita,Natarajan,Akhlaghi,Renema,Zhang12} and 
we evaluate the fidelity between the experimentally reconstructed POVM 
of our measurement device and the ideal SCS measurement in Eq.~(\ref{eq1}). 
We experimentally demonstrate the fidelities that cannot be achieved 
by conventional measurements such as homodyne detector 
or photon number resolving detector (PNRD).

Our scheme is inspired by the so-called quantum receiver idea in optical 
communication where the purpose of the receiver is to discriminate 
coherent states with minimum error probabilities \cite{Helstrom_book76_QDET}. 
The implementation of such receiver has been extensively explored in theory 
\cite{Kennedy73,Dolinar73,Bondurant93,TakeokaSasakiLutkenhaus2006_PRL_BinaryProjMmt,Takeoka2008,Guha2011,izumi2012,izumi2013} and experiment 
\cite{CookMartinGeremia2007_Nature,Wittmann2008_PRL_BPSK, Tsujino2010_OX_OnOff, Tsujino2011_Q_Receiver_BPSK,Mueller2012_NJP,Chen2012,Becerra13,Becerra15,Monteiro,Guerreiro,Izumiphase}. 
In this scenario, it is known that the optimal quantum measurement 
for discriminating the binary phase shift keyed (BPSK) coherent states 
$|{\pm}\alpha\rangle$ is given by Eq.~(\ref{eq1}) with specific sets of 
$\{c_0, c_1\}$ and $\phi=0$ (i.e. real superposition). 
Here we generalize such measurement by including arbitrary complex 
superposition ($\phi \ne 0$) and 
also by directly evaluating the structure of its POVM via QDT.

It is worth to mention the related work in Ref.\cite{Zhang12} 
where they performed a full detection tomography of a hybrid measurement of 
homodyne and PNRD in continuous variable Hilbert space and was able to reveal 
the wave-particle duality in measurement process. 
In contrast, the purpose of our work is to implement specific but nontrivial POVMs in the Hilbert space spanned by the SCS bases in  Eq.~(\ref{eq1}). 
Note that though it is two-dimensional, the SCS bases consist of the so-called continuous variable state vectors, and thus its implementation and tomographic verification are nontrivial. 
To do so, we develop a modified QDT technique that is of independent interest. 
Our technique has direct implications in applications using SCS states 
and their measurements such as 
quantum computation with optical coherent states \cite{Ralph,Lee}
or optimal detection of coherent states in optical communication 
\cite{Helstrom_book76_QDET,Kennedy73,Dolinar73,Bondurant93,TakeokaSasakiLutkenhaus2006_PRL_BinaryProjMmt,Takeoka2008,Guha2011,izumi2012,izumi2013,CookMartinGeremia2007_Nature,Wittmann2008_PRL_BPSK, Tsujino2010_OX_OnOff, Tsujino2011_Q_Receiver_BPSK,Mueller2012_NJP,Chen2012,Becerra13,Becerra15}, where the homodyne measurement and the photon counting are non-optimal.

\section{SCS measurement} 
\begin{figure}[t]
\centering
{
\includegraphics[width=0.95\linewidth]
{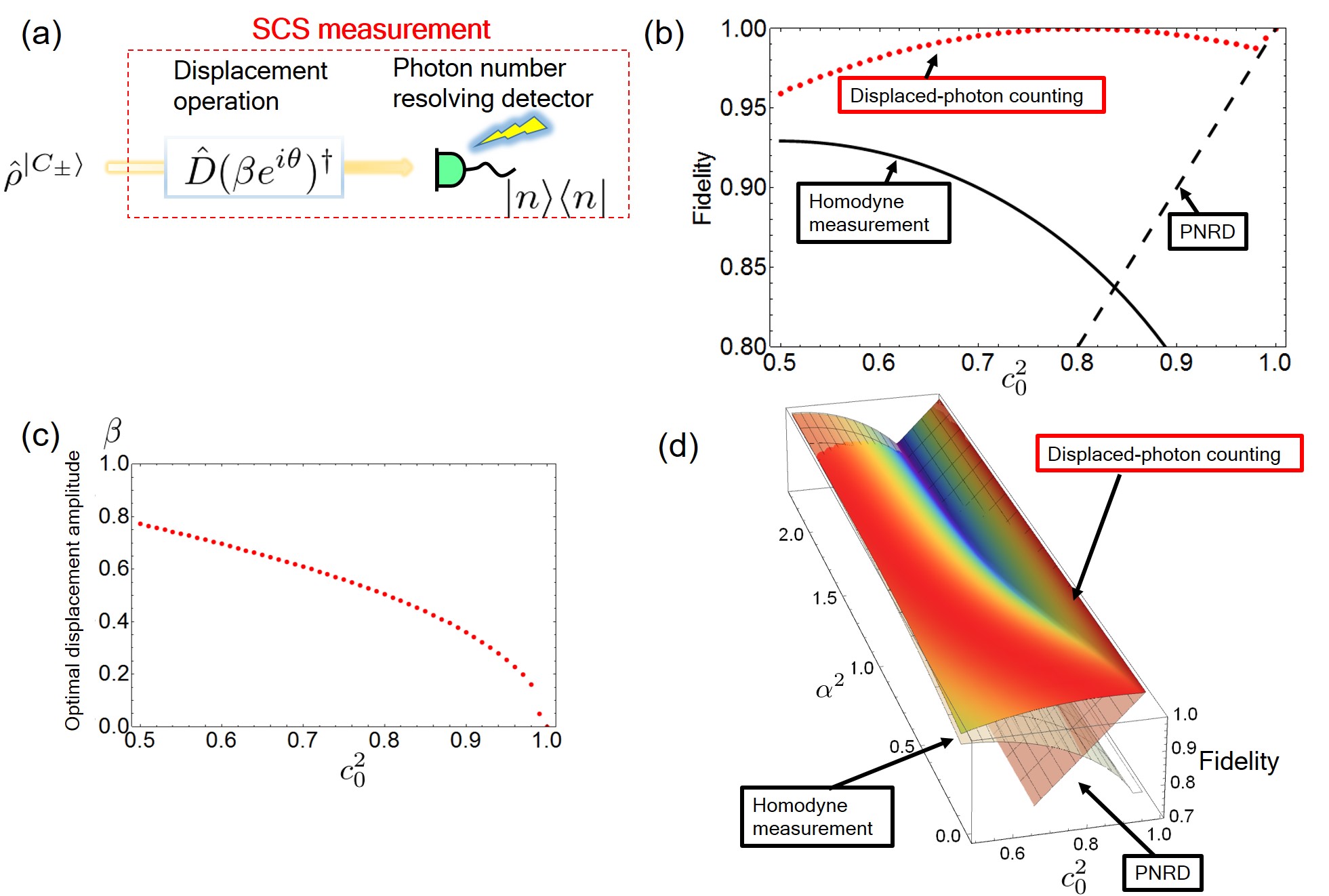}
}
\caption{
(a) Simple schematic of our measurement scheme. $\hat{\rho}^{\ket{C_{\pm}}}$ indicates the state that can be described in the two-dimensional space $\ket{C_{\pm}}$.
(b) Fidelity as a function of the superposition coefficient $c_0^2$ with fixed coherent amplitude $\alpha=0.50$ and phase $\phi=0$.
Dotted line, solid line and dashed line denote the displaced-photon counting,
the homodyne measurement and the PNRD respectively.
(c) Corresponding optimal displacement amplitude as a function of the superposition coefficient $c_0^2$.
(d) Fidelity as a function of the superposition coefficient $c_0^2$ and the mean photon number $\alpha^2$.The relative phase of the coefficients is set to $\phi=0$.
}
\label{Fidelity}
\end{figure}

Figure~\ref{Fidelity} (a) is a schematic of our measurement which approximately realizes the projection in Eq.~(\ref{eq1}).
It consists of a PNRD preceded by a displacement operation, which we call 
the displaced-photon counting hereafter. 
The measurement operators of our schematic in Fig.~\ref{Fidelity} (a) 
are given by,
\begin{eqnarray}
\hat{\Pi}_0^{\mathrm{DP}} & = & \sum_{\omega_0} \hat{D}(\beta) |n \rangle \langle n| \hat{D}(\beta )^{\dagger}
\nonumber\\ 
&&
\quad
\{ \omega_0 | \quad |\langle\pi_0^{SCS}| \hat{D}(\beta) |n\rangle|^2 
\ge |\langle\pi_1^{SCS}| \hat{D}(\beta) |n\rangle|^2  \}, 
\nonumber
\\
\hat{\Pi}_1^{\mathrm{DP}} & = & \hat{I} - \hat{\Pi}_0^{\mathrm{DP}}.
\label{eq2}
\end{eqnarray}
where the displacement operation $\hat{D}(\beta)=\exp[(\beta \hat{a}^{\dagger} - \beta^{\ast} \hat{a})] $ allows us to flexibly modulate the amplitude and phase of the coherent state $\ket{\gamma}$ such that $\hat{D}(\beta) \ket{\gamma} = \ket{\gamma+\beta}$.
The displacement operation is physically implemented by combining the signal state with a local oscillator (LO) at a beam splitter with nearly unit transmittance. 
A measurement operator of the PNRD is given by a set of photon number bases $\{\hat{\Pi}_n =\ket{n} \bra{n} \}$.

The intuition explaining how Eq. (\ref{eq2}) approximates Eq.~(\ref{eq1}) is as follows.
If the coherent amplitude is small, the SCS bases defined in the Eq.~(\ref{eq1}) can also be simply tailored 
in a superposition of a vacuum and single photon bases,
\begin{eqnarray}
|\pi^{SCS}_0\rangle &=& c_0 |C_+\rangle + c_1 \mathrm{e}^{i \phi} |C_-\rangle
\nonumber
\\
&\propto&
(c_0 / \mathcal{N}_+)\ket{0} +(c_1 \mathrm{e}^{i \phi} / \mathcal{N}_-)\alpha \ket{1}+\cdots.
\end{eqnarray}
Similarly the POVM in Eq.~(\ref{eq2}) is approximated by 
$\hat{\Pi}^{\rm DP}_0 = \hat{D}(\beta)|0\rangle\langle0|\hat{D}(\beta)$ 
and its complement for $\hat{\Pi}^{\rm DP}_1$. Then we observe that 
\begin{equation}
\hat{D}(\beta)\ket{0} = \ket{\beta} \propto \ket{0} +\beta \ket{1}+\cdots.
\end{equation}
The coefficient of the single photon basis in the above equation can be freely controlled by adjusting the amplitude and phase of the displacement operation. 
Thus the combination of the displacement operation and the photon counting provides high fidelity with the SCS measurement in small coherent amplitude region.
The amplitude and phase of the displacement operation are numerically optimized so as to maximize the fidelity between the SCS measurement and the displaced-photon counting measurement that is defined as,
\begin{equation}
\mathcal{F}^{\mathrm{DP}}= ( \langle\pi_0^{SCS}| \hat{\Pi}_0^{\mathrm{DP}} |\pi_0^{SCS}\rangle 
+ \langle\pi_1^{SCS}| \hat{\Pi}_1^{\mathrm{DP}} |\pi_1^{SCS}\rangle ) / 2.
\label{eq3}
\end{equation}

To evaluate the performance of our measurement strategy compared with the conventional measurements, 
we calculate the fidelities of the SCS measurement with the homodyne measurement and the PNRD without displacement.
We define the POVM of the homodyne measurement with binary outcomes as $\{ \hat{\Pi}_0^{\mathrm{HD}} = \int _{x_{\mathrm{th}}}^{\infty} \ket{x_{\phi}}\bra{x_{\phi}} d x_{\phi}, \hat{\Pi}_1^{\mathrm{HD}} = \hat{I} - \hat{\Pi}_0^{\mathrm{HD}} \}$, where   $\{ \ket{x_{\phi}} \}$ is the quadrature basis and $\phi$ is adjustable by changing the optical phase of the local oscillator.
We determine the threshold value $x_{\mathrm{th}}$ such that the fidelity 
\begin{equation}
\mathcal{F}^{\mathrm{HD}}= ( \langle\pi_0^{SCS}| \hat{\Pi}_0^{\mathrm{HD}} |\pi_0^{SCS}\rangle 
+ \langle\pi_1^{SCS}| \hat{\Pi}_1^{\mathrm{HD}} |\pi_1^{SCS}\rangle ) / 2 ,
\label{eq5}
\end{equation}
is maximized.
The PNRD is given by setting $\beta=0$ in Eq.~(\ref{eq2}). 
The fidelity to the SCS is then given by a simple form,
\begin{eqnarray}
\mathcal{F}^{\mathrm{PN}}=\left\{ \begin{array}{ll}
c_0^2 & (0.5\leq c_0^2 \leq 1) \\
c_1^2 & (0\leq c_0^2 < 0.5) 
\end{array} \right.
\label{eq6}
\end{eqnarray}

\begin{figure*}[t]
\centering
{
\includegraphics[width=0.9 \linewidth]{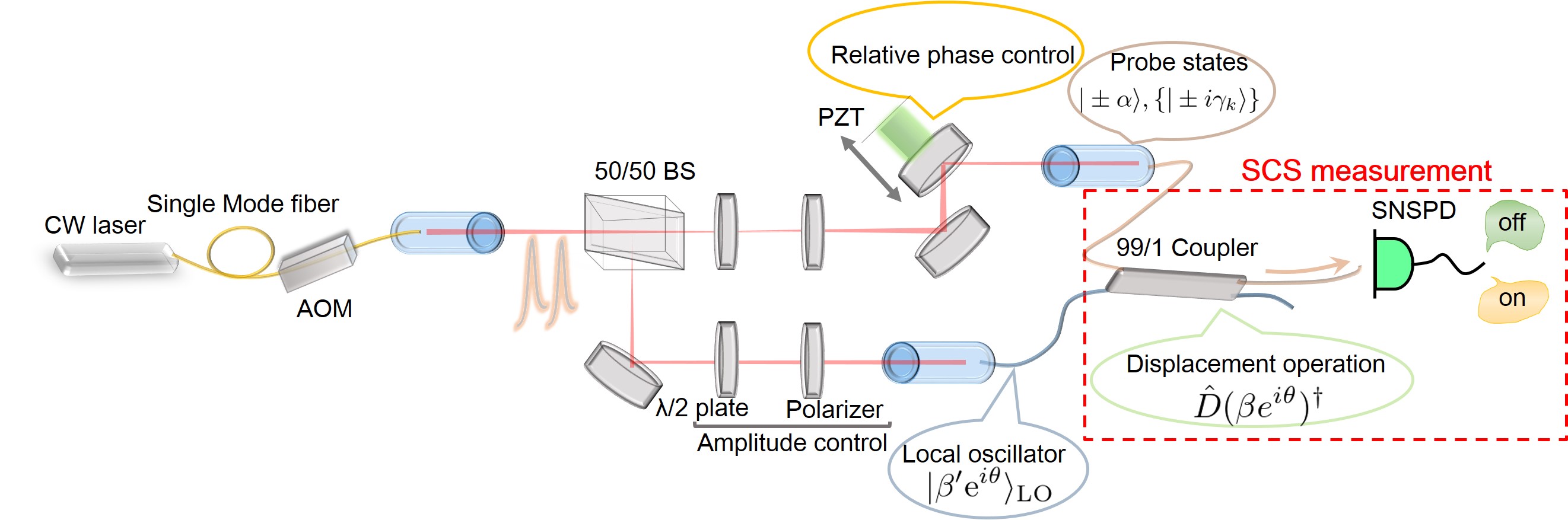}
}
\caption{
Experimental setup. 
AOM : acousto-optic modulator, 
BS :  beam splitter, 
PZT : piezo transducer,
SNSPD : superconducting nanowire single photon detector. 
The measurement process framed by red dashed line corresponds to the SCS measurement.
}
\label{setup}
\end{figure*}
We compare the fidelities for the three-type of measurements in Fig.~\ref{Fidelity} (b).
The relative phase and the coherent amplitude of the target SCS measurement are set to $\phi=0$ and $\alpha=0.50$ respectively.
The displaced-photon counting shows high fidelity for a whole range of $c_0$.
Figure~\ref{Fidelity} (c) depicts the optimal displacement amplitude as a function of the superposition coefficient $c_0^2$.
The optimal amplitude of displacement is decreased with the increase of the target coefficient and reaches to zero at $c_0^2=1$ where the SCS measurement can be achieved by the parity measurement using the PNRD.
Also, as will be shown later, our scheme can approximate the SCS measurement 
with a complex phase factor ($\phi\neq 0, \pi$) by optimizing both amplitude and phase of the displacement.
In Fig~\ref{Fidelity} (d) we evaluate the fidelities as a function of the superposition coefficient $c_0^2$ ($0.5\leq c_0^2\leq 1.0$) and the square of the coherent amplitude $\alpha^2$ ($0.1\leq\alpha^2\leq2.3$).
The displaced-photon counting offers a clear advantage over the conventional measurements up to $\alpha^2 \approx 1.5.$
A possible approach to achieve high fidelity for arbitrary $\{c_0, c_1, \phi, \alpha \}$ will be addressed in Discussion.
\section{Experiment} 

Our experimental setup is depicted in Fig.~\ref{setup}.
We prepare a sequence of optical pulses at a telecom wavelength 1549nm with repetition rate 900kHz and pulse width 100ns by modulating continuous wave laser using an acousto-optic modulator (AOM).
The optical pulse is first divided into two parts where one is 
the local oscillator for the displacement and the other is the probe pulse 
for the measurement characterization. 
For each state, we adjust the optical amplitude independently by means of a set of a half wave plate and a polarizer.
The probe state is interfered with the LO light on an asymmetric fiber coupler with transmittance $\tau=0.99$, which leads to the physical implementation of the displacement operation, and detected by the photon counter.
We achieve
the visibility  = 0.998 for the displacement operation.
In the experiment, instead of the PNRD, we use a superconducting nanowire 
single photon detector (SNSPD) which is capable of  
discriminating if the photon exists (on) or not (off) \cite{SSPD, SSPD2}.

The degradation of the fidelity due to the lack of the photon number resolution is negligible when the coherent amplitude $\alpha$ is small enough 
such that the probability of having more than one photon is negligible. 
Detection efficiency and dark count noise of the SNSPD are experimentally measured to be $68.9\%$ and $5.32\times 10^{-5}$ counts per pulse respectively.
The optical relative phase between the probe state and the local oscillator for the displacement operation, which determines the phase $\phi$ of the SCS measurement, is controlled by a piezo transducer (PZT).
We acquire $2\times10^5$ 
 experimental data for each probe state.

The implemented POVM is experimentally characterized by quantum detector 
tomography. 
Characterization over the entire phase space is in principle possible 
by sweeping the quadrature amplitudes of coherent state 
\cite{Lundeen09,Zhang12}. 
In the continuous variable QDT, usually one has to prepare a set of different quadrature amplitudes that entirely cover the phase space. 
However, here we can drastically reduce the number of quadrature amplitudes since our measurement device is on the space spanned by ${\ket{C_{\pm}}}$ which is intrinsically two-dimensional. 
Nevertheless, it is still not an easy task to tomograph it since each basis is a highly nonclassical continuous variable state.
For a two-dimensional space, four different probes 
are enough for the tomography, in our case
$\ket{\pm \alpha}$ and $(\ket{\alpha}\pm i\ket{-\alpha})/\sqrt{2}$. 
The coherent states $\ket{\pm \alpha}$ are easy to prepare. 
In contrast, to prepare well-calibrated SCSs as the probe is still challenging 
with the current technology. 
Thus we develop a method which replaces the SCS probes by $2k$-set of coherent states $\{ \ket{\pm i \gamma_{k}} \}$ with various amplitudes. 
Details of the method is discussed in Method.  
By numerically simulating the proposed QDT method, we find that the  
four probe states $\{\ket{\pm i\gamma_1},\ket{\pm i\gamma_2}\}$ and the coherent probes $\ket{\pm \alpha}$ suffice to characterize our measurement.
The set of the probe states  $\{\ket{\pm \alpha},\ket{\pm i\gamma_1},\ket{\pm i\gamma_2}\}$ and their measurement outcomes enable us to 
reconstruct the POVM and
we adopt the maximally likelihood procedure for the reconstruction \cite{Fiurasek}.
\begin{figure}[t]
\centering
{
\includegraphics[width=0.8\linewidth]
{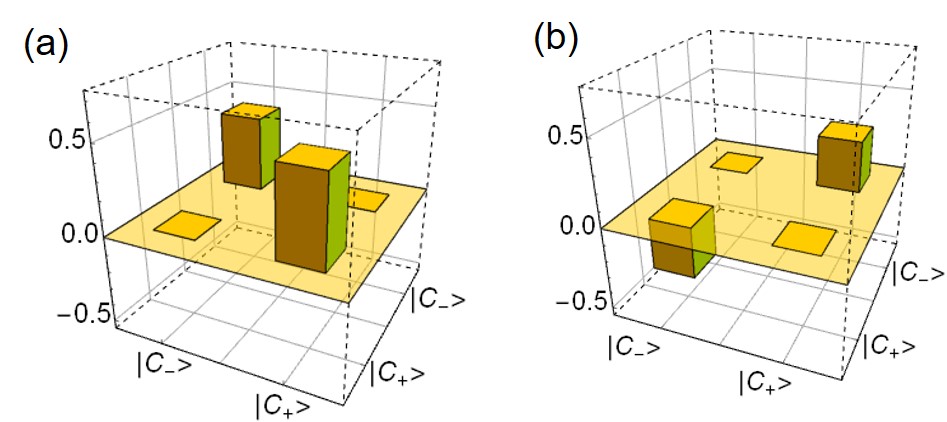}
}
\caption{ 
(a) Real and (b) imaginary part of the reconstructed POVM.
The reconstructed POVM elements are (a) $[[0.839\pm 0.000, 0.000\pm 0.000], [0.000\pm 0.000, 0.362\pm 0.013]]$ and (b) $[[0.000\pm 0.000, -0.237\pm 0.001], [0.237\pm 0.001, 0.000\pm0.000]]$.
The displacement amplitude is set to $\beta=0.894$ and the phase of the displacement with respect to the probe state $\ket{\alpha}$ is fixed to $\pi/2$.
}
\label{Result}
\end{figure}

\begin{figure}[h]
\centering
{
\includegraphics[width=0.95\linewidth]
{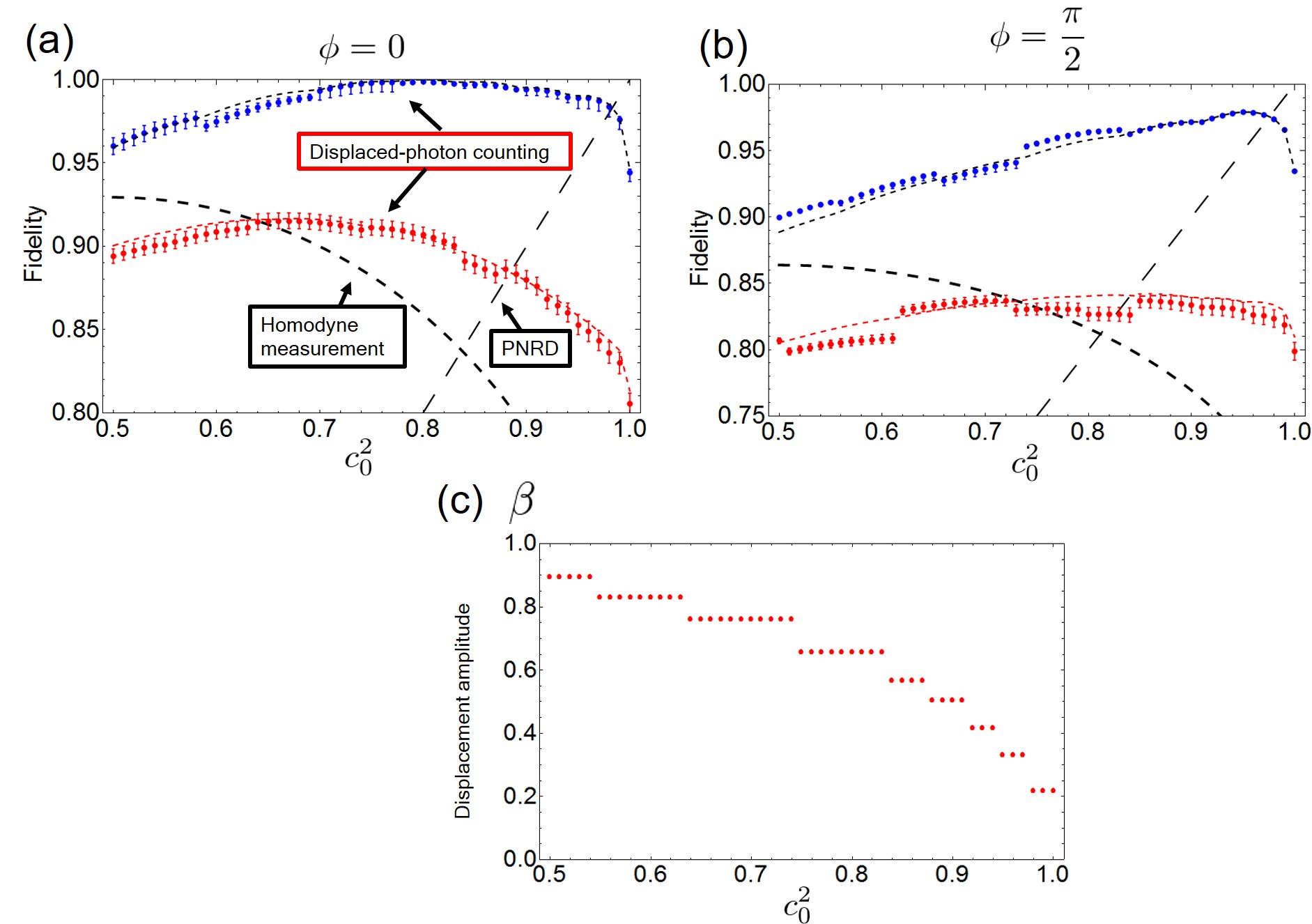}
}
\caption{
Fidelity as a function of the superposition coefficient $c_0^2$ with the coherent amplitude $\alpha=0.499$ and the relative phase, 
(a) $\phi=0$ and (b) $\phi=\pi/2$.
The displaced photon counting in the ideal case (blue dashed line),
the homodyne measurement (black dashed line), the PNRD (black long-dashed line), the displaced photon counting 
in the experimental condition for theory (red dashed line) and experiment (red circles), the experimental results with loss compensation (blue circles) are shown.
(c) Experimentally prepared amplitude of the displacement operation for the result (a).
}

\label{result1}
\end{figure}

An example of the experimentally reconstructed POVM is depicted 
in Fig.~\ref{Result}.
The amplitude and the phase of the target SCS bases are $\alpha=0.499$ and $\phi=\pi/2$ respectively.
As a corresponding measurement,
we prepare the local oscillator for the displacement operation with the amplitude $\beta=0.894$ and the relative phase $\pi/2$ with respect to $\ket{\alpha}$. 
Figure ~\ref{result1} (a) and (b) plot the fidelities between the target SCS measurement 
and the experimentally reconstructed displaced-photon counting 
for various $c_0^2$ (red circles). 
The blue circles are the same plots after compensating the loss. 
These plots are compared with their theoretical curves (red and blue dashed lines) 
and the theoretical curves for the ideal homodyne (black dashed line) and 
PNRD (black long-dashed line) measurements. 

In Fig.\ref{result1} (a) and (b), the target SCS amplitude and the relative phase are
$\alpha=0.499$
and, (a) $\phi=0$ and (b) $\phi=\pi/2$, respectively. 
Experimental results indicate that we can realize the SCS measurements with the fidelity better than both the ideal homodyne measurement and the ideal PNRD in the specific $c_0^2$ range.
Furthermore, by compensating the loss due to non-unit detection efficiency,  our experimental results outperform the ideal homodyne and the ideal PNRD in a whole range of $c_0^2$ except 
 $c_0^2 \sim 1.0$ where the photon number resolving capability is required.
As shown in Fig.~\ref{Fidelity} (c), the optimal amplitude of the displacement operation varies depending on the coefficient of the target SCS measurement.
We use 9 different displacement amplitudes shown in Fig.~\ref{result1} (c) to acquire the experimental data for Fig.~\ref{result1} (a) and the displacement amplitudes are chosen so as to maximize the fidelity under the experimental condition with finite loss. Therefore, the displacement amplitudes in Fig.~\ref{result1} (c) are slightly larger than those in Fig.~\ref{Fidelity} (c).
The effect of this stepwise displacement modulation appears 
in the discontinuity of the fidelity plots in Fig.~\ref{result1} (a).  
While the optimal fidelity is not obtainable with the stepwise displacement, the discrepancy of the fidelities between the optimal displacement and the experimental displacement condition is less than 0.2$\%$ except for $c_0^2=1.0$, where the optimal displacement amplitude is $\beta=0$, and the degradation of the fidelity due to non-optimal displacement is negligibly small.

A discrepancy between the theoretical prediction and the experimentally obtained fidelity can be explained as follows.
Red and blue dashed lines in Fig.~\ref{result1} (a) and (b) represent the theoretical fidelity between the displaced-photon counting and the SCS measurement with the coherent amplitude $\alpha=0.499$,
where $\alpha=0.499$ is determined by averaging the probe amplitude used to characterize each displacement condition.
The probe amplitude cannot be calibrated to the exactly same value due to the technical reasons and the systematic error of the probe amplitude is estimated to $\alpha=0.499\pm0.011$.
The error bar in Fig.~\ref{Result} and  Fig.~\ref{result1} is evaluated base on the systematic error of the probe amplitude.
In addition, the phase of the probe states with respect to the LO cannot be perfectly set to the desired value.
Both the finite precision of the amplitude and the phase make the experimental results higher or lower than the theoretical values.

\begin{figure}[h]
\centering
{
\includegraphics[width=1.0\linewidth]{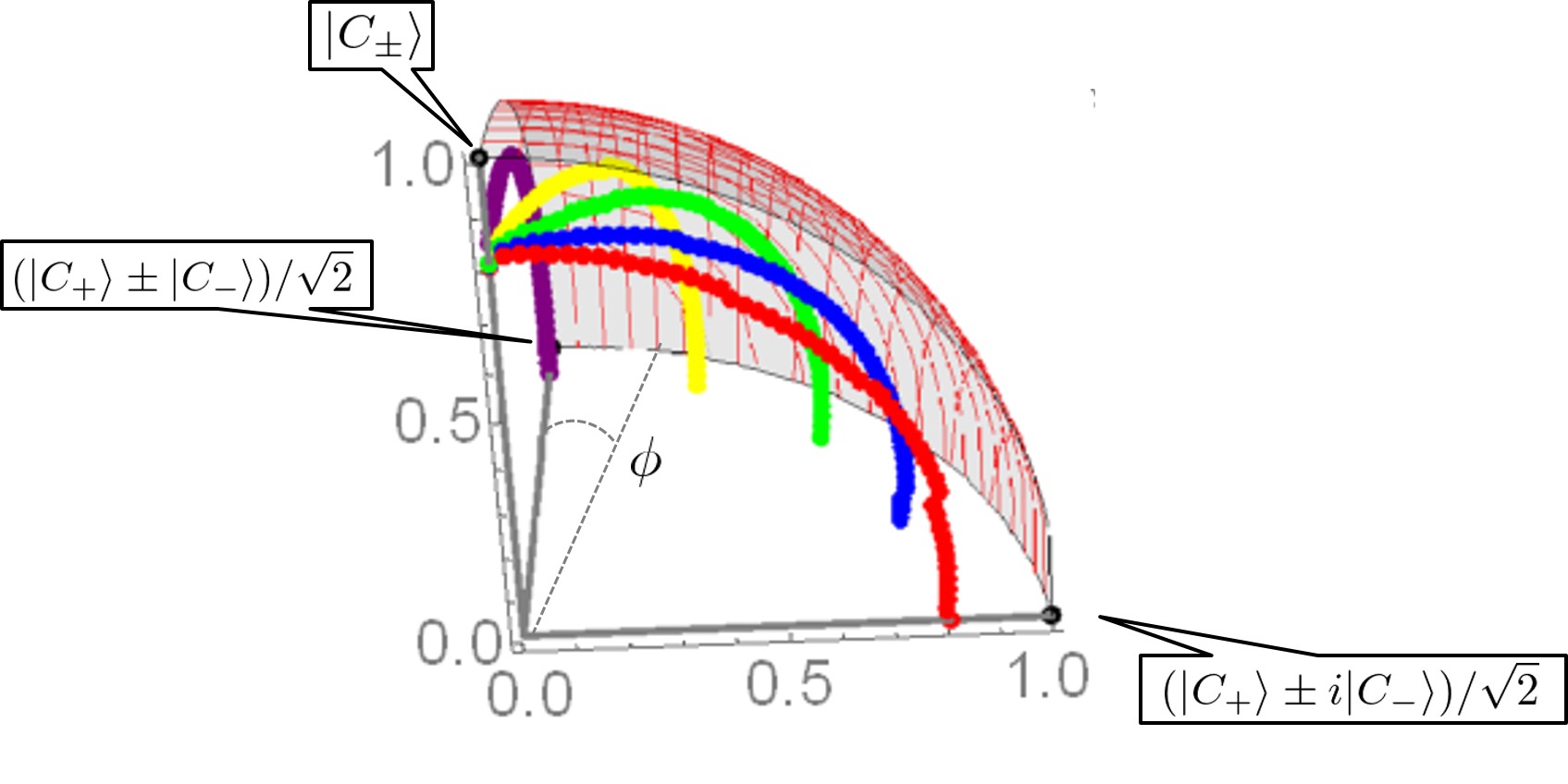}
}
\caption{
Experimentally obtained fidelities between the ideal SCS measurements and the displacement operation with the on-off counter.
We set the amplitude of the SCS and the phase to $\alpha=0.499$ and
$\phi=0$,
$0.393$,
$0.787$,
$1.18$,
$\pi/2$ from back to front.
}
\label{result2}
\end{figure}

Figure~\ref{result2} depicts a quadrant of a sphere with radius 1 in which experimentally obtained fidelities for various $\phi$ and $c_0$ are plotted.
The distance from the sphere origin to the plotted point corresponds to the fidelity between the target SCS and experimentally realized POVM.
Rotations in horizontal and vertical plane are equivalent to variation of $\phi$ and $c_0$ respectively.
We examine 5 different phase conditions $\phi=0, 0.393, 0.787, 1.18, \pi/2$ with the coherent amplitude $\alpha=0.499$ and the experimental results show that the arbitrary SCS measurement with weak coherent amplitude is approximately implementable by controlling both amplitude and phase of the displacement operation.

\section{Conclusion}
In this paper, we propose and experimentally demonstrated the physical implementation of the projection measurement onto the SCS bases.
Our theoretical analysis showed that the measurement process consisting of the displacement operation with the photon counter enables us to perform 
the SCS measurement with arbitrary $\{c_0, c_1, \phi\}$ in weak coherent amplitude case.
We demonstrated the proof-of-principle experiment for the SCS projection measurement and characterized our measurement by the QDT approach.
Although the fidelity between the ideal SCS measurement and the experimentally realized measurement was highly degraded because of the detector's imperfections, 
our experimental result showed higher fidelity than
the ideal homodyne measurement and the ideal PNRD for specific range of $c_0$.
Furthermore, by optimizing the amplitude and the phase of the displacement operation, we experimentally realized the approximate SCS measurement with arbitrary $\{c_0, c_1, \phi\}$.

An interesting future direction is the physical realization of the projection measurement onto the SCS bases with higher $\alpha$.
In fact it was shown that arbitrary two-dimensional projection measurement is achievable by introducing feedback operation to the displaced-photon counting measurement \cite{TakeokaSasakiLutkenhaus2006_PRL_BinaryProjMmt,Takeoka2008}.
The measurement strategy, which is often referred to as Dolinar receiver, was first proposed for the BPSK discrimination \cite{Dolinar73} and generalized for arbitrary two orthogonal optical states discrimination problem \cite{TakeokaSasakiLutkenhaus2006_PRL_BinaryProjMmt}.
Thus the displaced-photon counting with the feedback operation allows us to perform perfect SCS measurement with large coherent amplitude.
Another possible future work is the implementation of the SCS measurement for general input states.
Our analysis is concentrated on the two-dimensional space spanned by the SCS bases.
However, in principle,
it is also possible to realize the SCS measurement in higher dimensional space.
Such measurement procedure has not been explored but could be important tool for the optical QIP and communication scenarios.

\appendix

%
\section{Tomographic reconstruction of the displaced-photon counting 
in the SCS bases.}
%

The POVM of the displaced-photon counting is reconstructed by probing with coherent states 
and applying the QDT method. 
In general, the QDT requires a large amount of probe states to 
cover a whole Hilbert space of interest. 
In our case, though $|C_\pm\rangle$ is a continuous variable optical state, 
the signal space we are interested in is restricted to the two-dimensional space 
spanned by $\{ |C_+\rangle, |C_-\rangle \}$. 
Generally, the POVM tomography in a two-dimensional space requires at least 
four linearly independent probe states in the space \cite{QSEbook}. 
In our case, while the real part of the POVM 
is easily probed via two coherent states $|{\pm}\alpha\rangle$,  
it is necessary to use a superposition of $|{\pm}\alpha\rangle$ 
with imaginary phase to probe the imaginary part of the POVM. 
An example of such a state is 
\begin{equation}
\label{eq:3rd_probe}
|\phi_{\rm Im}^{+}\rangle = \frac{1}{\sqrt{2}} \left( 
|\alpha\rangle + i |{-}\alpha\rangle \right) ,
\end{equation}
which is not available in the laboratory with enough quality at present. 
Instead, we develop a method with the use of extra coherent states 
with imaginary valued amplitudes $|i\gamma_k\rangle$ to obtain 
the probing statistics 
for $|\phi_{\rm Im}^{+}\rangle$. 

In the following, we describe the method to reconstruct 
$\langle \phi_{\rm Im}^{+} | \hat{\Pi}_j | \phi_{\rm Im}^{+} \rangle$ 
without the SCS states.
 $\{\hat{\Pi}_j\}_j$ is the POVM to be reconstructed
and its representation in the photon number basis is
\begin{equation}
\label{eq:POVM_Fock}
\hat{\Pi}_j \equiv \sum_{m,n} \theta_{mn}^{(j)} 
|m \rangle\langle n| .
\end{equation}
If the probe $|\phi_{\rm Im}^{+}\rangle$ is available, 
its expectation value is given as 
\begin{eqnarray}
\label{eq:phi_im_expectation}
\langle \phi_{\rm Im}^{+} | \hat{\Pi}_j | \phi_{\rm Im}^{+} \rangle 
& = &
\frac{1}{2} \left\{ \langle\alpha| \hat{\Pi}_j |\alpha\rangle 
+ \langle{-}\alpha| \hat{\Pi}_j|{-}\alpha\rangle \right.
\nonumber\\ && \left.
+ i \left( \langle\alpha| \hat{\Pi}_j |{-}\alpha\rangle 
- \langle{-}\alpha| \hat{\Pi}_j|\alpha\rangle \right) \right\} .
\nonumber\\
\end{eqnarray}
The first two terms can be obtained
by using the probe states $|{\pm}\alpha\rangle$.
The last two terms are expressed as 
\begin{eqnarray}
\label{eq:POVM_third_fourth}
&& i \left( \langle\alpha|\hat{\Pi}_j|{-}\alpha\rangle 
- \langle{-}\alpha|\hat{\Pi}_j|\alpha\rangle \right)
\nonumber\\ 
&& = -i e^{-|\alpha|^2} 
\sum_{m,n} \frac{\alpha^{m+n}}{\sqrt{m!n!}} 
\left\{ (-1)^m - (-1)^n \right\} \theta^{(j)}_{mn} ,
\nonumber\\ 
&& = e^{-|\alpha|^2} 
\left\{ 2\alpha \Theta_{01}^{(j)}  + 2\alpha^3 (\Theta_{03}^{(j)}  - \Theta_{12}^{(j)} ) \right.
\nonumber\\ 
&& \left. \quad
+ 2\alpha^5 (\Theta_{05}^{(j)}  - \Theta_{14}^{(j)}  + \Theta_{23}^{(j)} ) + \cdots \right\} 
\nonumber\\ 
&& = 2 e^{-|\alpha|^2} 
\left( \alpha \Phi^{(j)}_1 + \alpha^3 \Phi^{(j)}_3 +  \alpha^5 \Phi^{(j)}_5 + \cdots \right) ,
\end{eqnarray}
where $i \Theta_{mn}^{(j)}  \equiv (\theta_{mn}^{(j)}  - \theta_{nm}^{(j)} )/\sqrt{m!n!}$ and 
$\Phi_1 = \Theta_{01}$, $\Phi_3 = \Theta_{03} - \Theta_{12}$, 
$\Phi_5 = \Theta_{05} - \Theta_{14} + \Theta_{23}$, etc. 
Note that $\Phi_l$ is always real. 
The following restriction on $\Phi^{(l)}$ is applied
by assuming that experimentally obtained POVMs are always physical, i.e., 
$\hat{\Pi}_j$ is positive definite and $\sum_j \hat{\Pi}_j = \hat{I}$. 
\begin{eqnarray}
\label{eq:Phi}
|\Phi_l| & = & \left| \sum_{m+n=l} \frac{\theta_{mn}}{2\sqrt{m!n!}} \left(
(-1)^n - (-1)^m \right) \right|
\nonumber\\
& = & 
\sum_{m=1}^l \frac{{\rm Im}\theta_{m(l-m)}}{2\sqrt{m!(l-m)!}} \left(
(-1)^m - (-1)^{(l-m)} \right) 
\nonumber\\ 
& \le &
\sum_{m=1}^l \frac{1}{\sqrt{m!(l-m)!}} .
\end{eqnarray}
See the next section for the proof of the inequality in the last line. 
Interestingly, the expectation values for the coherent state probes 
$|{\pm} i \gamma_k\rangle$ are also given 
as a function of $\Phi^{(j)}$,  
\begin{eqnarray}
\label{eq:gamma_expectation}
&& \langle i\gamma_k| \hat{\Pi}_j |i\gamma_k \rangle 
- 
\langle {-}i\gamma_k| \hat{\Pi}_j |{-}i\gamma_k \rangle 
\nonumber\\
&& = 
e^{-|\gamma_k|^2} \sum_{m,n} \frac{(i\gamma_k)^{m+n}}{\sqrt{m!n!}} 
\left\{ (-1)^m - (-1)^n \right\} \theta_{mn}^{(j)}  
\nonumber\\ 
&& = 2 e^{-|\gamma_k|^2} 
\left( -\gamma_k \Phi^{(j)}_1 + \gamma_k^3 \Phi^{(j)}_3 
-  \gamma_k^5 \Phi^{(j)}_5 + \cdots \right) .
\end{eqnarray}
Therefore, we can obtain the third and fourth terms in
 Eq.~(\ref{eq:phi_im_expectation}) 
by first characterizing $\{\Phi_i^{(j)}\}$ from the experimental 
results with $|\pm i\gamma_k\rangle$ and then substituting 
them into Eq.~(\ref{eq:POVM_third_fourth}). 

The experimentally measured count rates corresponding to Eq.~(\ref{eq:gamma_expectation}) is described as,
\begin{equation}
\label{eq:f}
f_k^{(j)} = \frac{1}{2 e^{-|\gamma_k|^2}} \left( 
\frac{ f^{(j)} (i\gamma_k) }{ \sum_l f^{(l)} (i\gamma_k) }
- \frac{ f^{(j)} ({-}i\gamma_k) }{ \sum_l f^{(l)} ({-}i\gamma_k) }
\right) ,
\end{equation}
where
$f^{(j)} (\pm i\gamma)$ denotes
the experimentally measured statistic for $\hat{\Pi}_j$ with the probe $|\pm i\gamma\rangle$.
A set of $\Phi^{(j)}_l$ up to $l=2K-1$ 
can be obtained from the probes $\{ |i\gamma_k\rangle \}$ 
($k=1, \cdots , K$) 
by solving the optimization problem 
\begin{eqnarray}
\label{eq:optimization_problem}
&& {\rm min} \{ || f^{(j)} - \Gamma \Phi^{(j)} ||_2 \} 
\nonumber\\ &&
{\rm subject \, to} \sum_{m=1}^l \frac{1}{\sqrt{m! (l-m)!}} - |\Phi^{(j)}_l| \ge 0 ,
\end{eqnarray}
where $|| v ||_2 \equiv (\sum_i |v_i|^2)^{1/2}$ is vector norm, 
$f^{(j)} = [ f^{(j)}_1, f^{(j)}_2, \cdots , f^{(j)}_K ]^T$, 
$\Phi^{(j)} = [ \Phi^{(j)}_1 , \Phi^{(j)}_3 , \cdots , \Phi^{(j)}_{2K-1} ]^T$, 
and 
\begin{equation}
\label{eq:Gamma}
\Gamma = \left[
\begin{array}{cccc}
-\gamma_1 & \gamma_1^3 & \cdots & (-1)^K \gamma_1^{2K-1} \\
\vdots & \vdots & \ddots & \vdots \\
-\gamma_K & \gamma_K^3 & \cdots & (-1)^K \gamma_K^{2K-1} 
\end{array}
\right] .
\end{equation}
The expectation value of our measurement for the probe state $\ket{\phi_{\mathrm{Im}}}$ is indirectly investigated by the set of coherent states $\{ \ket{\pm \alpha}, \ket{\pm i\gamma_k}\}$.
In a similar manner, 
we can obtain the expectation value for a state $|\phi_{\rm Im}^{-}\rangle = \left( 
|\alpha\rangle - i |{-}\alpha\rangle \right)/\sqrt{2}$ as well.
We numerically simulated our reconstruction procedure and
found that the probe states $\{\ket{\pm i\gamma_1}, \ket{\pm i\gamma_2}\}$, where the amplitudes are set to $\gamma=0.20, 0.30$, in addition to $ \ket{\pm \alpha}$ are sufficient to characterize the measurement in our experimental condition for the target SCS amplitude $\alpha=0.50$.
We apply the maximally likelihood method for the reconstruction of our measurement with the knowledge of the states $\{ \ket{\pm \alpha}, \ket{\phi^{\pm}_{\rm Im}} \}$ and their measurement results obtained from above process.

\section{Proof of the inequality in (\ref{eq:Phi})}

Let $\hat{\Pi}_l \equiv \sum_{i,j}^d \theta_{ij}|i \rangle\langle j|$
be the $l$th element of a POVM in a $d$ dimensional Hilbert space. 
To prove the inequality in (\ref{eq:Phi}), it is sufficient to show that 
all entries in $\hat{\Pi}_l$ satisfy $|\theta_{ij}|<1$.

We define the eigenvalues of $\hat{\Pi}_l$ as $\{ \lambda_1, \cdots, \lambda_d \}$ with a general constraint $0 \le \lambda_i \le 1$ ($i=1,\cdots,d$) and
\begin{equation}
\label{eq:unitary}
\hat{U}= 
\left[ \bar{u}_1 \, \bar{u}_2 \, \cdots \, \bar{u}_d \right]
\end{equation}
as a unitary matrix in the same space, 
where $\bar{u}_i$ is an orthonormal vector in the $d$ dimensional space,
\begin{equation}
\label{eq:uvec}
\bar{u}_i = \left[
\begin{array}{c} 
u_{i1} \\ \vdots \\ \ u_{id} 
\end{array}
\right].
\end{equation}
Then $\hat{\Pi}$ is decomposed as 
\begin{eqnarray}
\label{eq:u_lambda_u}
\hat{\Pi}_l & = & 
\hat{U}^\dagger {\rm diag}\left[\lambda_1, \cdots, \lambda_d\right]
\hat{U}
\nonumber\\
& = & 
\left[ \bar{u}_1 \, \cdots \, \bar{u}_d \right]^\dagger 
\left[ \lambda\bar{u}_1  \, \cdots \, \lambda\bar{u}_d \right] 
\nonumber\\
& = & 
\left[
\begin{array}{ccc}
\langle \bar{u}_1 ,\, \lambda\bar{u}_1 \rangle & \cdots & 
\langle \bar{u}_1 ,\, \lambda\bar{u}_d \rangle \\
\vdots & \ddots & \vdots \\
\langle \bar{u}_d ,\, \lambda\bar{u}_1 \rangle & \cdots & 
\langle \bar{u}_d ,\, \lambda\bar{u}_d \rangle 
\end{array}
\right] ,
\end{eqnarray}
where 
\begin{equation}
\label{eq:lambda_u}
\lambda\bar{u}_i = \left[
\begin{array}{c} 
\lambda_1 u_{i1} \\ \vdots \\ \lambda_d u_{id} 
\end{array}
\right], 
\end{equation}
and $\langle \bar{u} ,\, \bar{v} \rangle$ 
is the inner product of $\bar{u}$ and $\bar{v}$. 
Due to the property of unitary matrix and $\lambda_i$ we observe 
\begin{equation}
\label{eq:u_inner}
\langle \bar{u}_i ,\, \bar{u}_i \rangle = 1 , 
\quad
\langle \lambda\bar{u}_i ,\, \lambda\bar{u}_i \rangle \le 1, 
\end{equation}
and taking the Cauchy-Schwarz inequality, we have 
\begin{eqnarray}
\label{eq:cauchy-schwarz}
| \langle \bar{u}_i ,\, \lambda \bar{u}_j \rangle | & \le & 
\langle \bar{u}_i ,\, \bar{u}_i \rangle^{1/2}
\langle \lambda\bar{u}_j ,\, \lambda\bar{u}_j \rangle^{1/2}
\nonumber\\ & \le & 1 ,
\end{eqnarray}
which implies that the absolute value of 
each entry in the matrix of Eq.~(\ref{eq:u_lambda_u}) 
is always smaller than 1, i.e., $|\theta_{ij}|<1$.


\end{document}